\documentclass[twocolumn,pre,nofootinbib,floatfix,superscriptaddress,showkeys]{revtex4}

\usepackage{dcolumn}
\usepackage{graphicx}
\usepackage{rotating}
\usepackage{amsmath,amsfonts,amssymb}
\usepackage{multirow}

\newcommand{\e}{\mathrm{e}}
\renewcommand{\d}{\mathrm{d}}

\newcommand{\kg}{\,\textrm{kg}}
\newcommand{\g}{\,\textrm{g}}

\newcommand{\remove}[1]{}

\begin{document}

\title{How large should whales be?}
\author{Aaron Clauset}
\email{aaron.clauset@colorado.edu}
\affiliation{Department of Computer Science, University of Colorado, Boulder CO, 80309 USA}
\affiliation{BioFrontiers Institute, University of Colorado, Boulder CO, 80303 USA}
\affiliation{Santa Fe Institute, Santa Fe NM, 87501 USA}


\begin{abstract}
The evolution and distribution of species body sizes for terrestrial mammals is well-explained by a macroevolutionary tradeoff between short-term selective advantages and long-term extinction risks from increased species body size, unfolding above the $2\g$ minimum size induced by thermoregulation in air. Here, we consider whether this same tradeoff, formalized as a constrained convection-reaction-diffusion system, can also explain the sizes of fully aquatic mammals, which have not previously been considered. By replacing the terrestrial minimum with a pelagic one, at roughly $7000\g$, the terrestrial mammal tradeoff model accurately predicts, with no tunable parameters, the observed body masses of all extant cetacean species, including the $175,\!000,\!000\g$ Blue Whale. This strong agreement between theory and data suggests that a universal macroevolutionary tradeoff governs body size evolution for all mammals, regardless of their habitat. The dramatic sizes of cetaceans can thus be attributed mainly to the increased convective heat loss is water, which shifts the species size distribution upward and pushes its right tail into ranges inaccessible to terrestrial mammals. Under this macroevolutionary tradeoff, the largest expected species occurs where the rate at which smaller-bodied species move up into large-bodied niches approximately equals the rate at which extinction removes them.
\end{abstract}

\maketitle

\section{Introduction}
Cetaceans include the largest animals ever to live, including the Blue Whale (\textit{Balaenoptera musculus}), which is nearly 30 times larger than an African elephant and twice as large as the largest sauropod. However, the reasons for their enormous sizes or the possibility of still larger animals remains unclear. A deeper understanding of the evolutionary mechanisms shaping cetacean sizes would shed light on the role of energetic constraints in limiting species sizes~\cite{mcnab:2009}, and the interaction of macroecological patterns~\cite{brown:1995} and macroevolutionary processes~\cite{stanley:1975} in the oceans. It may also shed light on how long-term trends in species mass~\cite{alroy:2000a,alroy:2000b}, e.g., Cope's rule, the empirically observed tendency for species masses to increase within a lineage over evolutionary time~\cite{stanley:1973,alroy:1998}, operate in marine environments.

\begin{figure}[t!]
\centering
\includegraphics[scale=0.49]{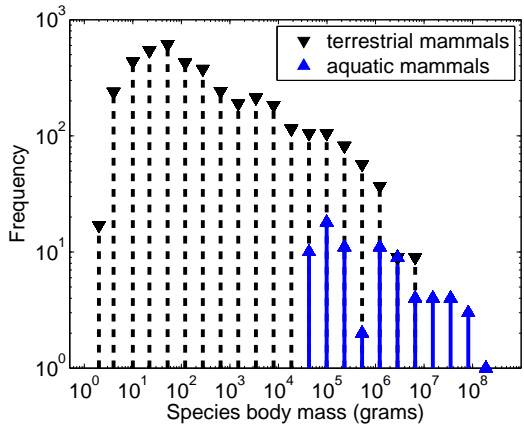}
\caption{\textbf{Terrestrial and fully aquatic mammal species mass distributions.} Both show the canonical asymmetric pattern: the median size is flanked by a short left-tail down to a minimum viable size and a long right-tail out to a few extremely large species.} 
\label{fig:sizes}
\end{figure}

Many major animal clades, including mammals, birds, fish and insects, seem to exhibit a canonical  pattern in the distribution of species masses~\cite{stanley:1973,kozlowski:gawelczyk:2002,allen:etal:2006,clauset:erwin:2008}. For example, the most common size of a terrestrial mammal is roughly $40\g$ (common Pacific Rat, {\em Rattus exulans}). Both larger and smaller species are much less common, but asymmetrically so: the largest species, like the extinct Imperial Mammoth ({\em Mammuthus imperator}, $10^{7}\g$), are orders of magnitude larger, while the smallest, like Remy's Pygmy Shrew ({\em Suncus remyi}, $2\g$), are only a little smaller (Fig.~\ref{fig:sizes}).

\begin{figure*}[t!]
\begin{tabular}{cc}
\includegraphics[scale=0.49]{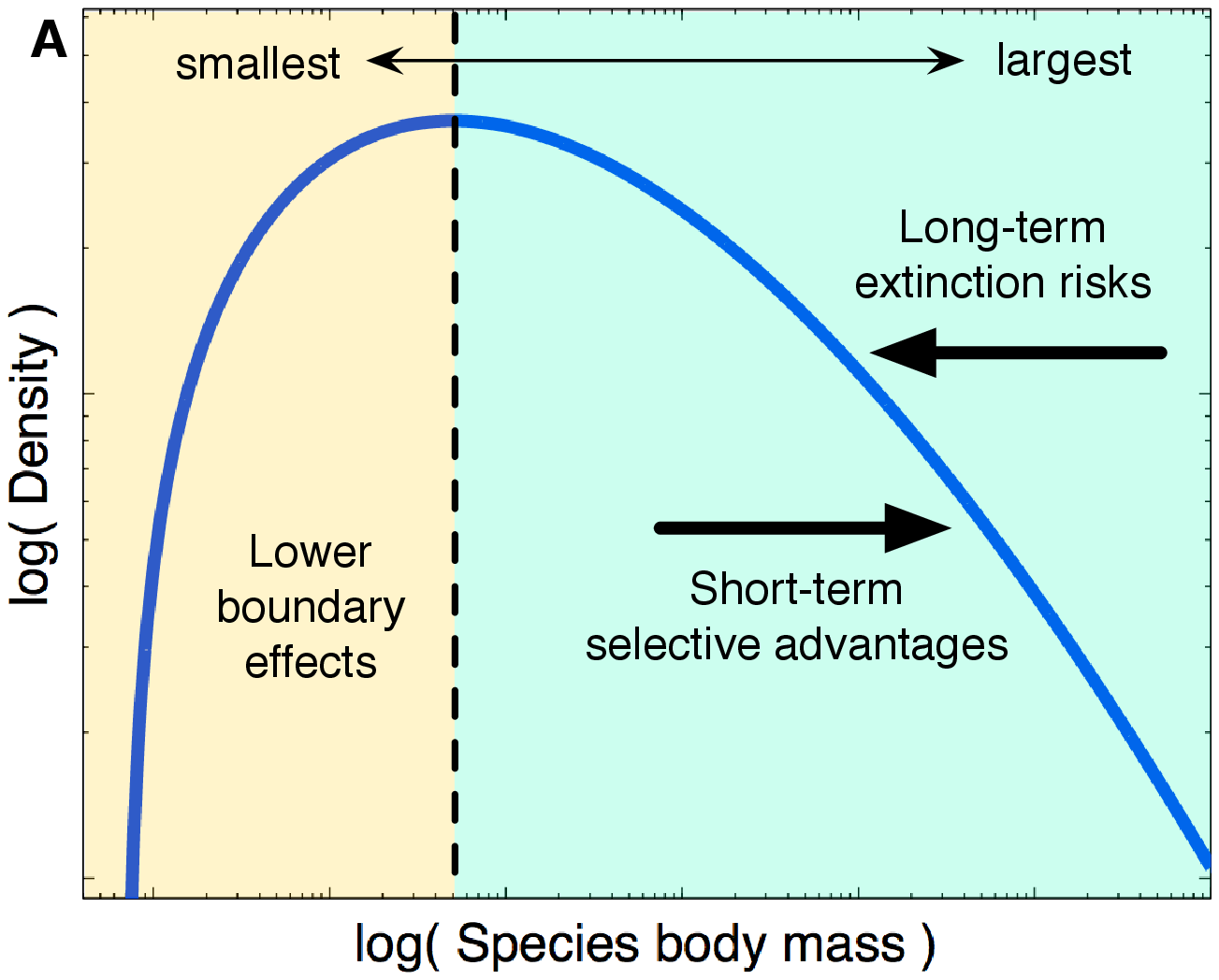} &
\includegraphics[scale=0.50]{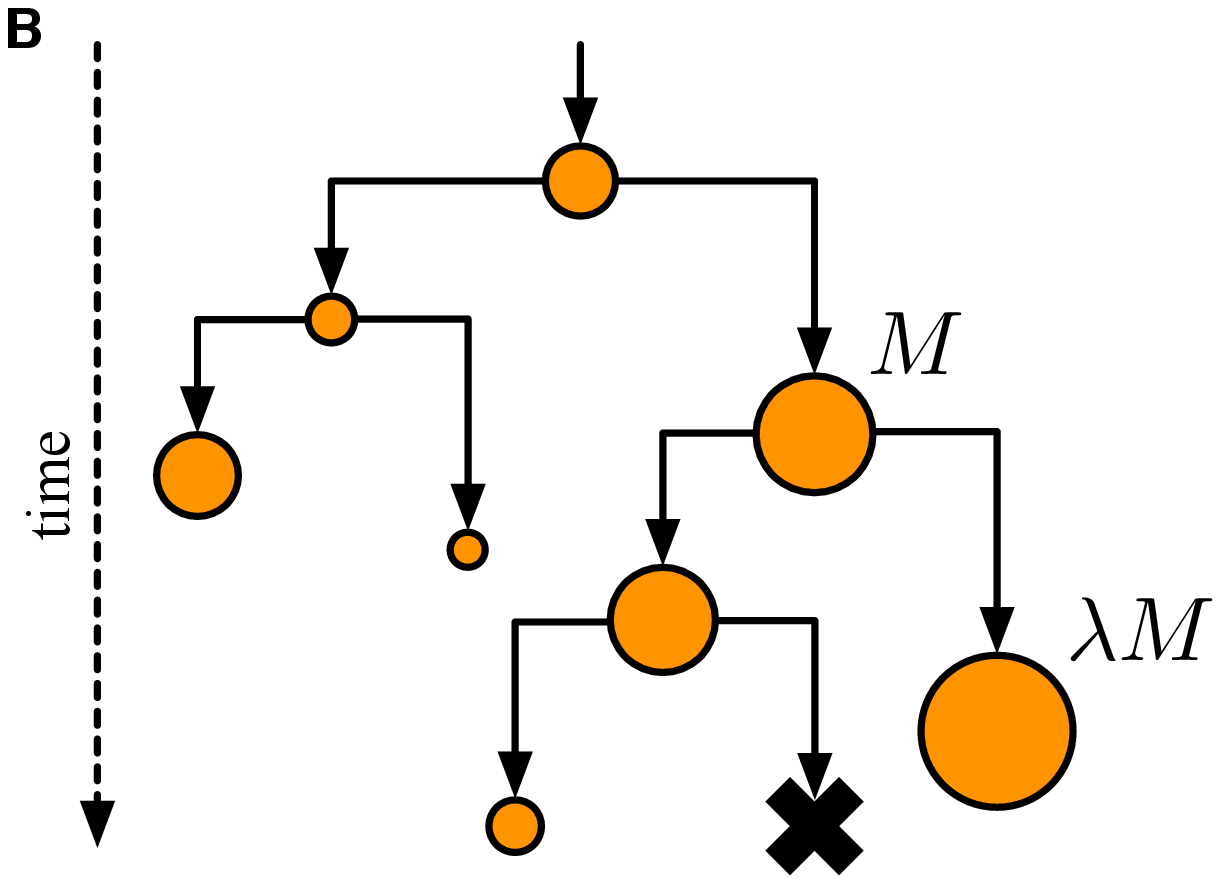}
\end{tabular} 
\caption{\textbf{Characteristic species size pattern and cladogenetic diffusion model.}
(A) The characteristic distribution of species body sizes, observed in most major animal groups. Macroevolutionary tradeoffs between short-term selective advantages and long-term extinction risks, constrained by a minimum viable size $M_{\min}$, produce the distribution's long right-tail. (B) Schematic illustrating the cladogenetic diffusion model of species body-size evolution: a descendant species' mass is related to its ancestor's size $M$ by a random multiplicative factor $\lambda$. Species become extinct with a probability that grows slowly with $M$.
} 
\label{fig:model}
\end{figure*}

Both the precise shape and the origins of this ubiquitous pattern have long been a topic of ecological~\cite{smith:lyons:2011} and evolutionary~\cite{stanley:1975,mcshea:1994} interest. 
Recently, this pattern was shown to be a long-term evolutionary consequence when a minimum viable body size, e.g., from physiological or thermoregulatory limits~\cite{ahlborn:2000}, constrains a macroevolutionary tradeoff between short-term selective advantages~\cite{brown:1995} and long-term extinction risks from increased species size~\cite{clauset:erwin:2008,clauset:redner:2009} (Fig.~\ref{fig:model}A). Early versions of this model~\cite{stanley:1975,mcshea:1994} demonstrated that species size evolution in the presence of a fixed lower limit produces right-skewed distributions that are qualitatively similar to the empirical pattern. However, these models also predict an unending increase in the size of the largest species, without necessarily adding new species. The key missing mechanism is extinction risk, which empirically tends to increase with species body size~\cite{liow:etal:2008,davidson:etal:2012} and thereby limit the number and size of large species. In this way, the characteristic pattern in species sizes can be explained from simple macroevolutionary mechanisms: speciation, variation, extinction and a physiological minimum size.

Historically, the main alternative explanation assumed the existence of a taxon-specific energetically optimal body size~\cite{lomolino:1985,sebens:1987,brown:etal:1996}. At this size, species maximize their ``reproductive power,'' i.e., the rate at which they convert environmental resources into offspring. Dispersion away from this optimum size was interpreted as evidence of interspecific competition. However, this theory remains controversial and, among other reasons~\cite{kozlowski:gawelczyk:2002}, contradicts strong evidence from the fossil record in the form of Cope's rule, a general statistical tendency for descendant species to be larger than their ancestors~\cite{alroy:1998,clauset:erwin:2008}, and the fact that most species are not close to their group's predicted optimal size.

Although the macroevolutionary tradeoff hypothesis has been quantitatively tested for extant terrestrial mammals~\cite{clauset:erwin:2008} and birds~\cite{clauset:etal:2009:a}, and its temporal dynamics have been shown to agree with the expansion of terrestrial mammals in the late Cretaceous and early Paleogene~\cite{clauset:redner:2009}, it remains unknown precisely how general this hypothesis is. For instance, it is unknown whether it holds for subclades of Mammalia, for fully aquatic mammals (which have typically been omitted from previous analyses), for ectothermic species, etc.

We resolve several of these questions by testing the tradeoff theory's ability to explain the observed body size distribution of cetaceans, the largest and most diverse marine mammal clade. Cetaceans are an ideal test case for the theory. First, Cetacea is a sufficiently speciose clade (77 extant species) to allow a quantitative comparison of predicted and observed distributions. Sirenia, the only other fully aquatic mammal clade, contains four extant species, which is too small for a productive comparison. Second, semiaquatic groups like Pinnipeds (seals and walruses) and Mustelids (otters) cannot be used to test the theory because they spend significant time on land, thus avoiding the hard thermoregulatory constraint assumed by the theory. Thus, by focusing on cetaceans, we provide a reasonable test of the theory. Third, fully aquatic mammals like cetaceans have typically been omitted in past studies because their marine habitat induces a different lower limit on mass than is seen in terrestrial mammals. As a result, it remains unknown whether the theory extends to all mammals, or only those in terrestrial environments. Finally, cetacean body masses do indeed exhibit the canonical right-skewed pattern (Fig.~\ref{fig:sizes}): the median size ($356\kg$, \textit{Tursiops truncatus}) is close to the smallest ($37.5\kg$, \textit{Pontoporia blainvillei}) but far from the largest ($175,\!000\kg$). This suggests that the theory may indeed hold for them.

Here, we test the strongest possible form of the macroevolutionary tradeoff theory for cetacean sizes. Instead of estimating model parameters from cetacean data, we combine parameters estimated from terrestrial mammals with a theoretically determined choice for the lower limit on cetacean species body mass. The resulting model has no tunable parameters by which to adjust its predicted distribution. In this way, we answer the question of how large a whale should be: if the predicted distribution agrees with the observed sizes, the same short-term versus long-term tradeoff that determines the sizes of terrestrial mammals also determines the sizes of whales.

We find that this zero-parameter model provides a highly accurate prediction of cetacean sizes. Thus, a single universal tradeoff mechanism appears to explain the body sizes of all mammal species, but this mechanism must obey the thermoregulatory limits imposed by the environment in which it unfolds. It is this one difference---thermoregulation in air for terrestrial mammals and in water for aquatic mammals---that explains the different locations of their respective body size distributions. Energetic constraints, while a popular historical explanation for sizes, seem to be only part of the puzzle for understanding the distribution of species sizes. Under this macroevolutionary mechanism, the size of the largest observed species is set by the tradeoff between the extinction probability at large sizes and the rate at which smaller species evolve to larger body masses, both of which may depend partly on energetic and ecological factors.

\vspace{-2mm}
\section{Methods} 
Following Clauset and Erwin~\cite{clauset:erwin:2008}, we model the tradeoff hypothesis as constrained cladogenetic diffusion, which includes only simple stochastic processes like speciation, extinction, size variation, and a minimum viable size. Deviations between this null model and the observed sizes of species can be interpreted as the effects of processes omitted from the model, e.g., interspecific competition, environmental effects, etc. We then compare this model's predictions to the observed sizes of all extant cetacean species.

\vspace{-1mm}
\subsection{Neutral model of species body sizes}
Under the constrained diffusion model, a species of mass $M$ produces descendant species with masses $\lambda M$ (Fig.~\ref{fig:model}B), where $\lambda$ is a random variable summarizing the contributions from all sources of short-term selective effects on size~\cite{stanley:1973,mcshea:1994}, including environmental gradients, interspecific competition and resource acquisition. For each speciation event, a new $\lambda$ is drawn independently from a fixed distribution $\Pr(\lambda)$. The interpretation of this model for variation in size down a lineage is that size-related short-term selection effects are uncorrelated across the clade. As a result, the distribution of sizes within the clade will evolve according to a diffusion process, and the trajectory of any particular lineage follows a kind of random walk~\cite{raup:1977,hunt:2007}. If the average size change between ancestors and descendants within a lineage is biased toward larger sizes (Cope's rule), we have $\langle\ln\lambda\rangle>0$~\cite{alroy:1998}. Anagenetic variation, or size variation between speciation events, need not be modeled separately as its impact may be absorbed into the $\lambda$ that describes the variation at the speciation event.

However, species may not take any size and thus the diffusion process is constrained. On the upper end, the probability of species extinction rises gently with increasing size~\cite{liow:etal:2008,davidson:etal:2012}. This size-dependency for extinction compactly summarize the systematic contributions from all sources to the overall extinction risk of larger-sized species, including larger energetic requirements~\cite{mcnab:2009}, smaller species abundance~\cite{white:etal:2007}, and longer generational times~\cite{martin:palumbi:1993}. The net effect is a soft upper limit on species sizes, rather than a hard upper limit like those derived from energetic constraints alone~\cite{mcnab:2009}. Given a particular extinction risk curve, the number and size of the very largest species is determined by a macroevolutionary balance between the upward ``pressure'' of smaller-sized lineages migrating into the larger size ranges~\cite{valkenburgh:etal:2004} and the downward extinction pressure of the increased extinction risk at those sizes.

On the lower end, endothermy imposes a minimum viable mass---a hard lower limit---that prohibits evolution toward ever smaller sizes. For terrestrial mammals and birds, this thermoregulatory minimum size is known to occur at roughly $M_{\min}=2\g$~\cite{pearson:1948,ahlborn:2000,west:etal:2002}, below which a species' convective heat loss in air is too high to maintain its internal temperature.

To extract a precise prediction of the species size distribution, we use a convection-diffusion-reaction formalization of the tradeoff theory~\cite{clauset:etal:2009:a,clauset:redner:2009}, which replaces the stochastic behavior of individual species and their lineages with a deterministic model of the relative density (fraction) of species at a given size. For analytic simplicity, we let the distribution of size changes $\Pr(\lambda)$ follow a log-normal distribution with parameters $v$ and $D$, an assumption that is consistent with fossil data~\cite{clauset:erwin:2008}.

Let $c(x,t)$ denote the density of species having mass $x=\ln M$ at time $t$. Under mild assumptions, the value $c(x,t)$ obeys the convection-diffusion-reaction equation in the continuum limit~\cite{berg:1993,krapivsky:etal:2010}:
\begin{equation}
\label{eq:cxt}
\frac{\partial c}{\partial t} +  v \frac{\partial c}{\partial x} = D \frac{\partial^2 c}{\partial x^2} + (k-A-Bx)c \enspace ,
\end{equation}
where $v=\langle\ln\lambda\rangle$ is the bias or average change in size from ancestor to descendent and $D=\langle (\ln\lambda)^2\rangle$ is the diffusion coefficient or the variance in size change. The expression $k-A$ is the size-independent (background) net speciation rate, which sets the absolute scale of the mass frequencies, and $B$ determines the strength and direction of a linear increase in extinction risk with the logarithm of species size.

In this model, the upper and lower size constraints guarantee the existence of a steady state distribution. To solve for its shape, we change variables $\mu=v/D$, $\alpha=(k-A)/D$, and $\beta=B/D$, and require that the distribution go to zero at $x=x_{\min}$. It can then be shown~\cite{clauset:etal:2009:a,clauset:redner:2009} that the steady-state distribution of sizes $x$ is
\begin{align}
c(x) \propto \e^{\mu\, x /2}{\rm Ai}\left[\beta^{1/3}(x-x_{\min})+z_{0}\right] \enspace , \label{eq:steady}
\end{align}
where ${\rm Ai}[.]$ is the Airy function and $z_{0}=-2.3381\dots$ is the location of its first zero. The shape of this curve is fully determined by three model parameters: $\mu$, the normalized strength of Cope's rule, $\beta$, the normalized size-dependence of extinction risk, and $x_{\min}$, the logarithm of the minimum viable body size. To compare the sizes predicted by these macroevolutionary processes with those observed in real species, we must only choose values for the model parameters.

For terrestrial mammals, estimates for $\mu$ and $\beta$ have previously been derived from fossil and extant data. The resulting size distribution accurately reproduces both the extant sizes of terrestrial mammals~\cite{clauset:erwin:2008} and their expansion during the late Cretaceous and early Paleogene~\cite{clauset:redner:2009,wilson:etal:2012}. Removing either the size-dependence of extinction risk or the minimum viable size produces unrealistic predictions~\cite{clauset:erwin:2008}.

The pelagic environments inhabited by cetaceans, however, impose distinct physiological, ecological and evolutionary challenges for endothermic mammals, and these are not reflected in the terrestrial model. One critical difference is the greater convective heat loss in water, which raises the minimum size of a competent aquatic endotherm. Thermoregulatory calculations and empirical data agree that this minimum size is roughly $M_{\min}=7\kg$~\cite{downhower:blumer:1988,ahlborn:blake:1999,ahlborn:2000}, about 3500 times larger than the minimum size imposed by thermoregulation in air.

\vspace{-1mm}
\subsection{Testing the tradeoff hypothesis}
A strong form of the macroevolutionary tradeoff hypothesis is to allow $M_{\min}$ to vary based on whether a species lives on land or in water, but to assume universal values for $\mu$ and $\beta$, i.e., values that hold regardless of habitat. By using estimates of $\mu$ and $\beta$ derived from terrestrial mammals alone, the model makes a prediction with no tunable parameters by which to adjust its fit to the observed cetacean sizes. This \textit{ex ante} prediction either matches the data or it does not.

To test the prediction, we constructed a novel body size data set covering all 77 extant cetacean species, from 183 empirical size estimates~\cite{long:1968,reeves:tracey:1980,mead:etal:1982,nagorsen:1985,stewart:stewart:1989,best:silva:1993,jefferson:etal:1993,jefferson:newcomer:1993,stacey:leatherwood:baird:1994,jefferson:leatherwood:1994:a,jefferson:leatherwood:1994:b,newcomer:jefferson:brownell:1996,jefferson:barros:1997,uhen:fordyce:barnes:1998:a,uhen:fordyce:barnes:1998:b,perrin:1998,clapham:mead:1999,cranford:1999,stacey:arnold:1999,jefferson:karczmarski:2001,perrin:2001,perrin:2002,smith:etal:2003,jefferson:curry:2003,perrin:zubtsova:kuzmin:2004,culik:2004,waerebeek:etal:2004,jefferson:hung:2004,borsa:2006}. Only plausibly independent, scientifically derived estimates were included. Mass ranges were converted to point estimates by taking their midpoint, unless a mean value was also provided. Subsequently, the mean value of all point estimates for a given species was used; this yielded an average of 2.4 measurements per species. Tables~\ref{table:1} and~\ref{table:2} give the mass estimates, primary source(s) and data curation comments.

We then evaluate the prediction's accuracy in two ways. First, we construct a classic hypothesis test for this ``zero parameter'' prediction. Such a test assumes observations are generated by independent draws from a fixed distribution, when in fact real species sizes are correlated due to shared evolutionary history. As a result, the hypothesis test is inherently conservative. Failure to reject the null model would indicate strong support for the tradeoff theory and that deviations between the predicted and observed size distributions are not statistically significant. Second, we consider whether the largest observed species, the Blue Whale, is statistically unlikely under the model. Failure to reject the hypothesis here indicates strong support for the number and size of very large species being set primarily by the macroevolutionary tradeoff, rather than by energetics alone.

\section{Results} \vspace{-2mm} 
Previous analyses of terrestrial mammal data~\cite{clauset:etal:2009:a,clauset:redner:2009} yielded $\mu\approx0.2$, a slight tendency toward larger sizes within a lineage (Cope's rule), and $\beta\approx0.08$, a weak tendency for extinction to increase with body size. Using these values and setting $M_{\min}=7\kg$ for fully aquatic species~\cite{downhower:blumer:1988} completes the model parameterization under Eq.~\eqref{eq:steady}.

\begin{figure}[t!]
\centering
\includegraphics[scale=0.49]{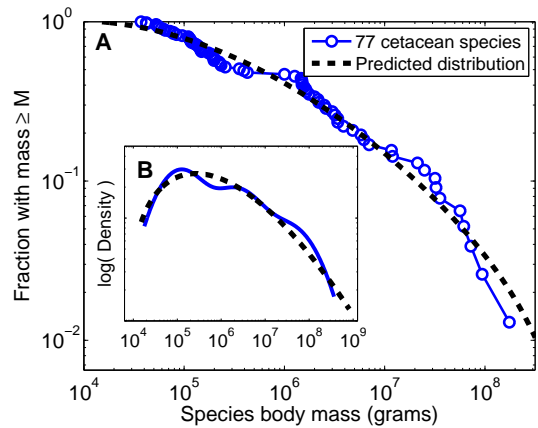}
\caption{\textbf{Comparison of data and model predictions.}
(A) \textit{Ex ante} predicted cetacean sizes, from a cladogenetic model fitted to terrestrial mammals but with a pelagic $M_{\min}$ (see text), and empirical sizes of 77 extant cetacean species, as complementary cumulative distributions and as (B) smoothed probability densities.
} 
\label{fig:prediction}
\end{figure}

Figure~\ref{fig:prediction} shows the predicted and observed distributions. The predicted model's statistical plausibility is determined by a standard two-tailed Kolmogorov-Smirnov hypothesis test, evaluated numerically. This produces a $p$-value of $p_{ks}=0.16\pm0.01$, which exceeds the conventional threshold for rejecting the null hypothesis. This indicates that the distribution of observed masses for cetacean species are statistically indistinguishable from the masses predicted by the model. As a control on the statistical uncertainty in the values of $\mu$ and $\beta$, we conduct a second test in which we add a small amount of Normally distributed noise to these parameter values and recompute $p_{ks}$ via Monte Carlo. This yields a slightly lower but still non-significant $p_{ks}=0.07\pm0.03$. 

We now consider whether the size of the largest observed cetacean species should be considered a statistical outlier under the model. The probability of observing at least one species with size at least as large as the Blue Whale at $M_{*}=1.75\times10^{8}\g$ was computed as $p(M_{*}) = 1-F(M_{*})^{n}$ where $F(M_{*})=\int_{M_{\rm min}}^{M_{*}}\Pr(M)\d M$ is the portion of the predicted distribution below $M_{*}$ and $n$ is the number of iid observations (extant species) drawn from $\Pr(M)$. Taking fixed parameters yields $p=0.91$, while simulating statistical uncertainty via Monte Carlo (as above) yields $p=0.88\pm0.03$, which is consistent with the fixed-parameter result.

These results imply that the observed sizes of whales are precisely what we would expect under a universal macroevolutionary tradeoff between short-term selective advantages and long-term extinction risks for increased size, unfolding under a constraint imposed by an environmentally-determined minimum viable size. This holds even for the enormous size of the Blue Whale, which is not statistically unlikely under this model. In fact, a species somewhat larger than the Blue Whale would also not be statistically unlikely, although no such species is known to have existed.

As a robustness check on our results, we test the assumption that $\beta$ takes a universal value for all mammals. Specifically, we hold $\mu$ fixed at the terrestrial value and estimate $\beta$ by fitting Eq.~\eqref{eq:steady} to the observed cetacean sizes. This procedure yields $\hat{\beta}_{\rm cete}\approx 0.097$, which is close to the terrestrial mammal value of $\beta\approx0.08$~\cite{clauset:etal:2009:a,clauset:redner:2009} and supports our assumption of a universal extinction risk curve for mammalian evolution. Furthermore, using this fitted value in the cetacean model, instead of the terrestrial value, would only reduce the statistical differences between the model and the data, and thus would not change our overall results.
A similar check on the universality of $\mu$ cannot be conducted at this time. The value of $\mu$ is most reliably estimated from comprehensive data on fossil species sizes~\cite{clauset:etal:2009:a}, which is not currently available for cetaceans.

\section{Discussion}
It is remarkable that the predicted distribution, which has no tunable parameters, is statistically indistinguishable from the observed sizes of cetaceans. Rarely in biological systems are the predictions of mathematical models so unambiguous and rarely are they upheld so clearly when compared to empirical data. This result thus strongly supports the hypothesis that both terrestrial and aquatic mammal sizes are shaped by a single universal macroevolutionary tradeoff between short-term advantages and long-term extinction risks of increased size, but which is constrained by a habitat-specific lower limit on size.

The only difference between our terrestrial and aquatic mammal tradeoff models is a larger minimum size for cetaceans, due greater convective heat loss in water. The macroevolutionary consequence is to shift upward the entire canonical species size distribution, pushing its right tail out into size ranges inaccessible to terrestrial mammals and producing giants like the Blue Whale. In this way we answer our motivating question of how large should whales be: they are as a group exactly as large as we should expect for mammals evolving under the thermoregulatory constraint of fully aquatic life. And, if we were given the first archaeocete's size, species counts over geological time and the model diffusion rate, the model would allow us to predict when a species of a given size should first have appeared.

The lower limit on size for a fully aquatic species would also have played a significant role over the long history of the Mammalia clade. From their emergence roughly 210~Ma to roughly 60~Ma, mammals were typically small-bodied~\cite{luo:2007}, with few or no species exceeding $M_{\min}$ for pelagic niches. Thus, aquatic lifestyles and the enormous body sizes associated with them would have been effectively inaccessible. In short, whales could not have evolved during this period. It was only in the late Cretaceous and early Paleogene, when the terrestrial mammal size distribution began expanding~\cite{clauset:redner:2009,wilson:etal:2012} that there were sufficient numbers of species above the threshold for a transition into pelagic habitats to be possible.

It is interesting to note that almost immediately after the terrestrial size distribution extended beyond the pelagic minimum, mammals did indeed invaded the oceans. This coincidence suggests a kind of body-size mediated ecological release, in which the expansion of the species size distribution enabled a dramatic and qualitative change in the large-scale occupation of ecological niches. For this reason, the historical timing of when mammals returned to the oceans is explained by the timing of late Cretaceous and early Paleogene expansion relative to the particular size required for a fully aquatic lifestyle.

On the upper end of sizes, some past work has considered the possibility of maximum species sizes due to energetic constraints~\cite{ahlborn:2000,mcnab:2009}. For instance, in the case of powered flight, decreasing metabolic power per unit mass effectively makes it difficult for birds above $15\kg$ to generate sufficient power for flapping flight~\cite{ahlborn:2000}. Of course, flightless birds like the ostrich (at roughly $100\kg$) have circumvented this constraint by abandoning flight altogether. In the case of cetaceans, recent work suggests a similar decreasing power delivery per unit mass during lunge feeding in large mysticetes~\cite{goldbogen:etal:2012,potvin:etal:2012}. This tendency suggests a maximum species size caused by the increased difficulty faced by very large whales in satisfying their energetic requirements. In principle, however, whales may be able to circumvent this limit by changing their feeding behavior or food source~\cite{potvin:etal:2012}.

Although it is reasonable to argue that whales cannot evolve to arbitrarily large sizes, it remains unclear whether a genuine maximum size from energetic constraints is low enough to impact the observed distribution of sizes. Our results suggest that there is no statistical evidence for such a limit in the vicinity of the Blue Whale's mass at $10^{8}\g$, as we achieve statistical indistinguishability without an explicit limit.
In fact, a slightly larger species would also not be statistically unlikely under the model, suggesting that the Blue Whale's size may arise more from its particular energetically-suboptimal lunging strategy~\cite{potvin:etal:2012} than from a fundamental limit on all possible cetaceans.

The macroevolutionary tradeoff theory does produce a general upper limit on size: the largest observed species occurs at a size close to where the net speciation rate effectively falls to zero, which is a finite value for any finite-sized clade. With $M_{\min}$ fixed by the environment, the precise location of this point depends on the rates at which smaller-bodied species evolve to larger sizes (captured by the model parameter $\mu$) and at which extinction eliminates them (captured by $\beta$). This type of macroevolutionary turnover at the largest sizes is known to have occurred repeatedly in North American canids~\cite{valkenburgh:etal:2004}. The tradeoff theory implies that the pattern is ubiquitous, and should also occur in cetaceans.

At the macroevolutionary level of analysis considered here, the effects of  energetics, population size, generation time, interspecific competition, morphology, geography, climate, etc.\ are all implicitly captured by the structure and parameters of the diffusion and extinction processes. The highly abstract nature of this theory does not undermine the importance of these factors for explaining the sizes of specific species in specific environments. It merely implies that across the clade and across large spatial and temporal scales, these factors collectively exert gentle macroevolutionary pressures that can be compactly summarized by a constrained diffusion model. For investigating species sizes within specific clades, the tradeoff hypothesis should be viewed as a kind of ``neutral'' model. Statistically significant deviations  imply the presence of non-neutral evolutionary or ecological processes. In the same way, changes in model parameter values over deep time may indicate broad-scale, non-stationary processes like climate change or clade-level ecological competition, as between mammals and dinosaurs prior to the K-Pg event.

In closing, we point out that this tradeoff between short-term advantages and long-term risks for increased size is entirely general. To date, however, it has only be tested on endotherms like mammals and birds. If the theory also holds for other major clades, such as aquatic tetrapod groups like icthyosaurs, plesiosaurs and turtles, groups like dinosaurs, fish and foraminifera, or subclades within these groups, it would have major implications for our understanding of macroevolution. A broad examination of minimum viable sizes and size-dependent extinction risks across groups and geologic time would thus better elucidate the role of these mechanisms in shaping the trajectory of species sizes throughout the history of life.

\begin{acknowledgments} 
The author thanks Doug Erwin and Sid Redner for helpful conversations. This work was supported in part by the Santa Fe Institute.
\end{acknowledgments}

\renewcommand{\thefigure}{S\arabic{figure}}
\setcounter{figure}{0}
\renewcommand{\thetable}{S\arabic{table}}
\setcounter{table}{0}

\newpage
\begin{table}
{\tiny 
\begin{tabular}{lll|r|p{4cm}|p{5.5cm}}
group   &   family   &   species                                    &   mass (kg)   &   primary source (reference)   &   curation comments  \\ 
\hline
  Mysticeti    &   Balaenidae   &   Balaena mysticetus   &   100000   &   Smith etal (2003)   &   .  \\ 
  Mysticeti    &   Balaenidae   &   Balaena mysticetus   &   87500   &   Jefferson Leatherwood Webber (1993)   &   .  \\ 
  Mysticeti    &   Balaenidae   &   Eubalaena australis   &   23000   &   Smith etal (2003)   &   .  \\ 
  Mysticeti    &   Balaenidae   &   Eubalaena australis   &   100000   &   Jefferson Leatherwood Webber (1993)   &   .  \\ 
  Mysticeti    &   Balaenidae   &   Eubalaena glacialis   &   23000   &   Smith etal (2003)   &   .  \\ 
  Mysticeti    &   Balaenidae   &   Eubalaena glacialis   &   90000   &   Jefferson Leatherwood Webber (1993)   &   .  \\ 
  Mysticeti    &   Balaenopteridae   &   Balaenoptera acutorostrata   &   10000   &   Smith etal (2003)   &   .  \\ 
  Mysticeti    &   Balaenopteridae   &   Balaenoptera acutorostrata   &   14000   &   Jefferson Leatherwood Webber (1993), Perrin Zubtsova Kuzmin (2004)   &   .  \\ 
  Mysticeti    &   Balaenopteridae   &   Balaenoptera borealis   &   20000   &   Smith etal (2003)   &   .  \\ 
  Mysticeti    &   Balaenopteridae   &   Balaenoptera borealis   &   30000   &   Jefferson Leatherwood Webber (1993), Long 1968   &   .  \\ 
  Mysticeti    &   Balaenopteridae   &   Balaenoptera edeni   &   20000   &   Smith etal (2003)   &   .  \\ 
  Mysticeti    &   Balaenopteridae   &   Balaenoptera edeni   &   22500   &   Jefferson Leatherwood Webber (1993)   &   .  \\ 
  Mysticeti    &   Balaenopteridae   &   Balaenoptera musculus   &   190000   &   Smith etal (2003)   &   .  \\ 
  Mysticeti    &   Balaenopteridae   &   Balaenoptera musculus   &   160000   &   Jefferson Leatherwood Webber (1993), Morton ed 1997   &   .  \\ 
  Mysticeti    &   Balaenopteridae   &   Balaenoptera physalus   &   70000   &   Smith etal (2003)   &   .  \\ 
  Mysticeti    &   Balaenopteridae   &   Balaenoptera physalus   &   75000   &   Jefferson Leatherwood Webber (1993), Uhen Fordyce Barnes 1998 inJanisGunnellUhen   &   .  \\ 
  Mysticeti    &   Balaenopteridae   &   Megaptera novaeangliae   &   30000   &   Smith etal (2003)   &   .  \\ 
  Mysticeti    &   Balaenopteridae   &   Megaptera novaeangliae   &   35000   &   Jefferson Leatherwood Webber (1993), Clapham Mead 1999, Uhen Fordyce Barnes 1998 inJanisGunnellUhen   &   .  \\ 
  Mysticeti    &   Eschrichtiidae   &   Eschrichtius robustus   &   28500   &   Smith etal (2003)   &   .  \\ 
  Mysticeti    &   Eschrichtiidae   &   Eschrichtius robustus   &   35000   &   Jefferson Leatherwood Webber (1993), Uhen Fordyce Barnes 1998 inJanisGunnellUhen   &   .  \\ 
  Mysticeti    &   Neobalaenidae   &   Caperea marginata   &   3200   &   Jefferson Leatherwood Webber (1993), Perrin Zubtsova Kuzmin (2004)   &   estimate by Smith etal 2003 gives a mass 10x as large, so we omit Smith etal  \\ 
  Odontoceti   &   Delphinidae   &   Cephalorhynchus commersonii   &   72.4   &   Smith etal (2003)   &   .  \\ 
  Odontoceti   &   Delphinidae   &   Cephalorhynchus commersonii   &   76   &   Jefferson Leatherwood Webber (1993)   &   .  \\ 
  Odontoceti   &   Delphinidae   &   Cephalorhynchus commersonii   &   86   &   Culik (2004)   &   .  \\ 
  Odontoceti   &   Delphinidae   &   Cephalorhynchus eutropia   &   45   &   Smith etal (2003)   &   .  \\ 
  Odontoceti   &   Delphinidae   &   Cephalorhynchus eutropia   &   63   &   Jefferson Leatherwood Webber (1993)   &   .  \\ 
  Odontoceti   &   Delphinidae   &   Cephalorhynchus eutropia   &   60   &   Culik (2004)   &   .  \\ 
  Odontoceti   &   Delphinidae   &   Cephalorhynchus heavisidii   &   40   &   Smith etal (2003)   &   .  \\ 
  Odontoceti   &   Delphinidae   &   Cephalorhynchus heavisidii   &   65   &   Culik (2004)   &   .  \\ 
  Odontoceti   &   Delphinidae   &   Cephalorhynchus hectori   &   50   &   Smith etal (2003)   &   .  \\ 
  Odontoceti   &   Delphinidae   &   Cephalorhynchus hectori   &   57   &   Jefferson Leatherwood Webber (1993)   &   .  \\ 
  Odontoceti   &   Delphinidae   &   Delphinus delphis   &   80   &   Smith etal (2003)   &   .  \\ 
  Odontoceti   &   Delphinidae   &   Delphinus delphis   &   135   &   Jefferson Leatherwood Webber (1993)   &   .  \\ 
  Odontoceti   &   Delphinidae   &   Delphinus delphis   &   200   &   Culik (2004)   &   .  \\ 
  Odontoceti   &   Delphinidae   &   Feresa attenuata   &   170   &   Smith etal (2003)   &   .  \\ 
  Odontoceti   &   Delphinidae   &   Feresa attenuata   &   225   &   Jefferson Leatherwood Webber (1993)   &   .  \\ 
  Odontoceti   &   Delphinidae   &   Globicephala macrorhynchus   &   726   &   Smith etal (2003)   &   .  \\ 
  Odontoceti   &   Delphinidae   &   Globicephala macrorhynchus   &   3600   &   Jefferson Leatherwood Webber (1993)   &   .  \\ 
  Odontoceti   &   Delphinidae   &   Globicephala melas   &   800   &   Smith etal (2003)   &   .  \\ 
  Odontoceti   &   Delphinidae   &   Globicephala melas   &   2000   &   Jefferson Leatherwood Webber (1993)   &   .  \\ 
  Odontoceti   &   Delphinidae   &   Globicephala melas   &   1600   &   Perrin Zubtsova Kuzmin (2004)   &   .  \\ 
  Odontoceti   &   Delphinidae   &   Grampus griseus   &   387.5   &   Smith etal (2003)   &   .  \\ 
  Odontoceti   &   Delphinidae   &   Grampus griseus   &   400   &   Jefferson Leatherwood Webber (1993)   &   .  \\ 
  Odontoceti   &   Delphinidae   &   Lagenodelphis hosei   &   164   &   Smith etal (2003)   &   .  \\ 
  Odontoceti   &   Delphinidae   &   Lagenodelphis hosei   &   210   &   Jefferson Leatherwood Webber (1993)   &   .  \\ 
  Odontoceti   &   Delphinidae   &   Lagenodelphis hosei   &   210   &   Culik (2004)   &   .  \\ 
  Odontoceti   &   Delphinidae   &   Lagenodelphis hosei   &   209   &   Jefferson Leatherwood 1994   &   .  \\ 
  Odontoceti   &   Delphinidae   &   Lagenorhynchus acutus   &   182   &   Smith etal (2003)   &   .  \\ 
  Odontoceti   &   Delphinidae   &   Lagenorhynchus acutus   &   208.5   &   Jefferson Leatherwood Webber (1993)   &   .  \\ 
  Odontoceti   &   Delphinidae   &   Lagenorhynchus acutus   &   205   &   Culik (2004)   &   .  \\ 
  Odontoceti   &   Delphinidae   &   Lagenorhynchus albirostris   &   180   &   Smith etal (2003)   &   .  \\ 
  Odontoceti   &   Delphinidae   &   Lagenorhynchus albirostris   &   265   &   Culik (2004)   &   .  \\ 
  Odontoceti   &   Delphinidae   &   Lagenorhynchus australis   &   120   &   Smith etal (2003)   &   .  \\ 
  Odontoceti   &   Delphinidae   &   Lagenorhynchus australis   &   115   &   Jefferson Leatherwood Webber (1993)   &   .  \\ 
  Odontoceti   &   Delphinidae   &   Lagenorhynchus australis   &   115   &   Culik (2004)   &   .  \\ 
  Odontoceti   &   Delphinidae   &   Lagenorhynchus cruciger   &   110   &   Smith etal (2003)   &   .  \\ 
  Odontoceti   &   Delphinidae   &   Lagenorhynchus obliquidens   &   120   &   Smith etal (2003)   &   .  \\ 
  Odontoceti   &   Delphinidae   &   Lagenorhynchus obliquidens   &   180   &   Jefferson Leatherwood Webber (1993)   &   .  \\ 
  Odontoceti   &   Delphinidae   &   Lagenorhynchus obliquidens   &   82.5   &   Culik (2004)   &   .  \\ 
  Odontoceti   &   Delphinidae   &   Lagenorhynchus obscurus   &   127.5   &   Smith etal (2003)   &   .  \\ 
  Odontoceti   &   Delphinidae   &   Lagenorhynchus obscurus   &   60   &   Jefferson Leatherwood Webber (1993)   &   .  \\ 
  Odontoceti   &   Delphinidae   &   Lagenorhynchus obscurus   &   100   &   Culik (2004)   &   .  \\ 
  Odontoceti   &   Delphinidae   &   Lissodelphis borealis   &   113   &   Smith etal (2003)   &   .  \\ 
  Odontoceti   &   Delphinidae   &   Lissodelphis borealis   &   115   &   Jefferson Leatherwood Webber (1993)   &   .  \\ 
  Odontoceti   &   Delphinidae   &   Lissodelphis borealis   &   116   &   Culik (2004)   &   .  \\ 
  Odontoceti   &   Delphinidae   &   Lissodelphis borealis   &   113   &   Jefferson Newcomer (1993)   &   .  \\ 
  Odontoceti   &   Delphinidae   &   Lissodelphis peronii   &   116   &   Smith etal (2003)   &   .  \\ 
  Odontoceti   &   Delphinidae   &   Lissodelphis peronii   &   116   &   Jefferson Leatherwood Webber (1993)   &   Culik 2004 repeats this measurement, so we omit Culik  \\ 
  Odontoceti   &   Delphinidae   &   Lissodelphis peronii   &   116   &   Newcomer Jefferson Brownell 1996   &   .  \\ 
  Odontoceti   &   Delphinidae   &   Orcaella brevirostris   &   190   &   Smith etal (2003)   &   .  \\ 
  Odontoceti   &   Delphinidae   &   Orcaella brevirostris   &   122.5   &   Culik (2004)   &   .  \\ 
  Odontoceti   &   Delphinidae   &   Orcaella brevirostris   &   123.5   &   Stacey Arnold 1999   &   .  \\ 
  Odontoceti   &   Delphinidae   &   Orcinus orca   &   4300   &   Smith etal (2003)   &   .  \\ 
  Odontoceti   &   Delphinidae   &   Orcinus orca   &   8750   &   Jefferson Leatherwood Webber (1993)   &   .  \\ 
  Odontoceti   &   Delphinidae   &   Orcinus orca   &   4685   &   Culik (2004)   &   .  \\ 
  Odontoceti   &   Delphinidae   &   Orcinus orca   &   7050   &   Perrin Zubtsova Kuzmin (2004)   &   reported mass mean of 2 specimens  \\ 
  Odontoceti   &   Delphinidae   &   Peponocephala electra   &   208   &   Smith etal (2003)   &   .  \\ 
  Odontoceti   &   Delphinidae   &   Peponocephala electra   &   275   &   Jefferson Leatherwood Webber (1993)   &   .  \\ 
  Odontoceti   &   Delphinidae   &   Peponocephala electra   &   228   &   Culik (2004)   &   .  \\ 
  Odontoceti   &   Delphinidae   &   Peponocephala electra   &   208   &   Jefferson Barros 1997   &   .  \\ 
  Odontoceti   &   Delphinidae   &   Pseudorca crassidens   &   1360   &   Smith etal (2003)   &   .  \\ 
  Odontoceti   &   Delphinidae   &   Pseudorca crassidens   &   2000   &   Jefferson Leatherwood Webber (1993)   &   .  \\ 
  Odontoceti   &   Delphinidae   &   Pseudorca crassidens   &   1360   &   Stacey Leatherwood 1994   &   .  \\ 
  Odontoceti   &   Delphinidae   &   Sotalia fluviatilis   &   44   &   Smith etal (2003)   &   .  \\ 
  Odontoceti   &   Delphinidae   &   Sotalia fluviatilis   &   40   &   Jefferson Leatherwood Webber (1993)   &   .  \\ 
  Odontoceti   &   Delphinidae   &   Sousa chinensis   &   265   &   Smith etal (2003)   &   .  \\ 
  Odontoceti   &   Delphinidae   &   Sousa chinensis   &   284   &   Jefferson Leatherwood Webber (1993)   &   .  \\ 
  Odontoceti   &   Delphinidae   &   Sousa chinensis   &   215   &   Culik (2004)   &   .  \\ 
  Odontoceti   &   Delphinidae   &   Sousa chinensis   &   265   &   Jefferson Karczmarksi (2001)   &   .  \\ 
  Odontoceti   &   Delphinidae   &   Sousa teuszii   &   100   &   Smith etal (2003)   &   .  \\ 
  Odontoceti   &   Delphinidae   &   Sousa teuszii   &   284   &   Jefferson Leatherwood Webber (1993)   &   .  \\ 
  Odontoceti   &   Delphinidae   &   Sousa teuszii   &   215   &   Culik (2004)   &   .  \\ 
  Odontoceti   &   Delphinidae   &   Sousa teuszii   &   166   &   Waerebeek etal (2004)   &   .  
\end{tabular}
}
\caption{Cetacean size estimates (part 1).}
\label{table:1}
\end{table}

\newpage

\begin{table}
{\tiny 
\begin{tabular}{lll|r|p{4cm}|p{5.5cm}}
group   &   family   &   species                                    &   mass (kg)   &   primary source (reference)   &   curation comments  \\ 
\hline
  Odontoceti   &   Delphinidae   &   Stenella attenuata   &   120   &   Jefferson Leatherwood Webber (1993)   &   estimate from Smith etal 2003 is less than 1/2 of the values reported elsewhere, so we omit Smith etal  \\ 
  Odontoceti   &   Delphinidae   &   Stenella attenuata   &   119   &   Culik (2004)   &   .  \\ 
  Odontoceti   &   Delphinidae   &   Stenella attenuata   &   119   &   Perrin (2001)   &   .  \\
  Odontoceti   &   Delphinidae   &   Stenella clymene   &   68   &   Smith etal (2003)   &   .  \\ 
  Odontoceti   &   Delphinidae   &   Stenella clymene   &   85   &   Jefferson Leatherwood Webber (1993)   &   .  \\ 
  Odontoceti   &   Delphinidae   &   Stenella clymene   &   80   &   Culik (2004)   &   .  \\ 
  Odontoceti   &   Delphinidae   &   Stenella clymene   &   80   &   Jefferson Curry (2003)   &   .  \\ 
  Odontoceti   &   Delphinidae   &   Stenella coeruleoalba   &   135.9   &   Smith etal (2003)   &   .  \\ 
  Odontoceti   &   Delphinidae   &   Stenella coeruleoalba   &   156   &   Jefferson Leatherwood Webber (1993)   &   .  \\ 
  Odontoceti   &   Delphinidae   &   Stenella coeruleoalba   &   156   &   Culik (2004)   &   .  \\
  Odontoceti   &   Delphinidae   &   Stenella frontalis   &   110   &   Smith etal (2003)   &   .  \\ 
  Odontoceti   &   Delphinidae   &   Stenella frontalis   &   143   &   Jefferson Leatherwood Webber (1993)   &   .  \\ 
  Odontoceti   &   Delphinidae   &   Stenella frontalis   &   143   &   Culik (2004)   &   .  \\ 
  Odontoceti   &   Delphinidae   &   Stenella frontalis   &   140   &   Perrin (2002)   &   .  \\ 
  Odontoceti   &   Delphinidae   &   Stenella longirostris   &   50.5   &   Smith etal (2003)   &   .  \\ 
  Odontoceti   &   Delphinidae   &   Stenella longirostris   &   77   &   Jefferson Leatherwood Webber (1993)   &   .  \\ 
  Odontoceti   &   Delphinidae   &   Stenella longirostris   &   50.5   &   Perrin 1998   &   Culik 2004 repeats measurement of Perrin 1998, so we omit Culik  \\
  Odontoceti   &   Delphinidae   &   Steno bredanensis   &   130   &   Smith etal (2003)   &   .  \\ 
  Odontoceti   &   Delphinidae   &   Steno bredanensis   &   150   &   Jefferson Leatherwood Webber (1993)   &   .  \\ 
  Odontoceti   &   Delphinidae   &   Steno bredanensis   &   155   &   Culik (2004)   &   .  \\ 
  Odontoceti   &   Delphinidae   &   Tursiops truncatus   &   175   &   Smith etal (2003)   &   .  \\ 
  Odontoceti   &   Delphinidae   &   Tursiops truncatus   &   650   &   Jefferson Leatherwood Webber (1993)   &   .  \\ 
  Odontoceti   &   Delphinidae   &   Tursiops truncatus   &   242   &   Culik (2004)   &   .  \\ 
  Odontoceti   &   Monodontidae   &   Delphinapterus leucas   &   1360   &   Smith etal (2003)   &   .  \\ 
  Odontoceti   &   Monodontidae   &   Delphinapterus leucas   &   1500   &   Steward Steward (1989), Uhen Fordyce Barnes 1998 inJanisGunnellUhen   &   .  \\ 
  Odontoceti   &   Monodontidae   &   Delphinapterus leucas   &   1500   &   Culik (2004)   &   .  \\ 
  Odontoceti   &   Monodontidae   &   Delphinapterus leucas   &   1600   &   Jefferson Leatherwood Webber (1993)   &   .  \\ 
  Odontoceti   &   Monodontidae   &   Monodon monoceros   &   900   &   Smith etal (2003)   &   .  \\ 
  Odontoceti   &   Monodontidae   &   Monodon monoceros   &   1600   &   Jefferson Leatherwood Webber (1993), Reeves Tracey (1980)   &   .  \\ 
  Odontoceti   &   Monodontidae   &   Monodon monoceros   &   1300   &   Culik (2004)   &   .  \\ 
  Odontoceti   &   Phocoenidae   &   Australophocaena dioptrica   &   65   &   Smith etal (2003)   &   .  \\ 
  Odontoceti   &   Phocoenidae   &   Neophocaena phocaenoides   &   32.5   &   Smith etal (2003)   &   .  \\ 
  Odontoceti   &   Phocoenidae   &   Neophocaena phocaenoides   &   85   &   Culik (2004)   &   .  \\ 
  Odontoceti   &   Phocoenidae   &   Neophocaena phocaenoides   &   71.8   &   Jefferson Hung (2004)   &   .  \\ 
  Odontoceti   &   Phocoenidae   &   Phocoena phocoena   &   52.5   &   Smith etal (2003)   &   .  \\ 
  Odontoceti   &   Phocoenidae   &   Phocoena phocoena   &   57.5   &   Jefferson Leatherwood Webber (1993)   &   .  \\ 
  Odontoceti   &   Phocoenidae   &   Phocoena phocoena   &   55   &   Culik (2004)   &   .  \\ 
  Odontoceti   &   Phocoenidae   &   Phocoena sinus   &   42.5   &   Smith etal (2003)   &   .  \\ 
  Odontoceti   &   Phocoenidae   &   Phocoena spinipinnis   &   60   &   Smith etal (2003)   &   .  \\ 
  Odontoceti   &   Phocoenidae   &   Phocoena spinipinnis   &   85   &   Jefferson Leatherwood Webber (1993)   &   .  \\ 
  Odontoceti   &   Phocoenidae   &   Phocoenoides dalli   &   102.5   &   Smith etal (2003)   &   .  \\ 
  Odontoceti   &   Phocoenidae   &   Phocoenoides dalli   &   200   &   Jefferson Leatherwood Webber (1993)   &   .  \\ 
  Odontoceti   &   Phocoenidae   &   Phocoenoides dalli   &   200   &   Culik (2004)   &   .  \\ 
  Odontoceti   &   Phocoenidae   &   Phocoenoides dalli   &   200   &   Jefferson (1988)   &   .  \\ 
  Odontoceti   &   Physeteridae   &   Kogia breviceps   &   431.5   &   Culik (2004), Borsa (2006), Uhen Fordyce Barnes (1998) inJanisGunnellUhen   &   .  \\ 
  Odontoceti   &   Physeteridae   &   Kogia breviceps   &   450   &   Culik (2004)   &   .  \\ 
  Odontoceti   &   Physeteridae   &   Kogia breviceps   &   400   &   Borsa (2006), Jefferson Leatherwood Webber (1993), Uhen Fordyce Barnes (1998) inJanisGunnellUhen   &   .  \\ 
  Odontoceti   &   Physeteridae   &   Kogia simus   &   183.1   &   Smith etal (2003)   &   .  \\ 
  Odontoceti   &   Physeteridae   &   Kogia simus   &   270   &   Nagorsen 1985   &   .  \\ 
  Odontoceti   &   Physeteridae   &   Kogia simus   &   270   &   Culik (2004)   &   mass quoted as 2702 kg, but this is too big by an order of magnitude. Assumed to be 270.2kg  \\ 
  Odontoceti   &   Physeteridae   &   Kogia simus   &   210   &   Jefferson Leatherwood Webber (1993)   &   .  \\ 
  Odontoceti   &   Physeteridae   &   Physeter catodon   &   14025   &   Smith etal (2003)   &   .  \\ 
  Odontoceti   &   Physeteridae   &   Physeter catodon   &   57000   &   Jefferson Leatherwood Webber (1993), Cranford (1999)   &   .  \\ 
  Odontoceti   &   Platanistidae   &   Inia geoffrensis   &   129.25   &   Smith etal (2003)   &   .  \\ 
  Odontoceti   &   Platanistidae   &   Inia geoffrensis   &   160   &   Jefferson Leatherwood Webber (1993)   &   .  \\ 
  Odontoceti   &   Platanistidae   &   Inia geoffrensis   &   167.5   &   Culik (2004)   &   .  \\ 
  Odontoceti   &   Platanistidae   &   Inia geoffrensis   &   129.25   &   Best Silva (1993)   &   .  \\ 
  Odontoceti   &   Platanistidae   &   Lipotes vexillifer   &   187.5   &   Jefferson Leatherwood Webber (1993)   &   mass estimate from Smith etal 2003 is less than 1/2 estimated range here, so we omit Smith etal  \\ 
  Odontoceti   &   Platanistidae   &   Platanista gangetica   &   115   &   Smith etal (2003)   &   .  \\ 
  Odontoceti   &   Platanistidae   &   Platanista gangetica   &   108   &   Jefferson Leatherwood Webber (1993)   &   .  \\ 
  Odontoceti   &   Platanistidae   &   Platanista minor   &   83.9146   &   Smith etal (2003)   &   .  \\ 
  Odontoceti   &   Platanistidae   &   Pontoporia blainvillei   &   40.5   &   Smith etal (2003)   &   .  \\ 
  Odontoceti   &   Platanistidae   &   Pontoporia blainvillei   &   34   &   Jefferson Leatherwood Webber (1993)   &   .  \\ 
  Odontoceti   &   Ziphiidae   &   Berardius arnuxii   &   7000   &   Smith etal (2003)   &   .  \\ 
  Odontoceti   &   Ziphiidae   &   Berardius bairdii   &   11380   &   Smith etal (2003)   &   .  \\ 
  Odontoceti   &   Ziphiidae   &   Berardius bairdii   &   12000   &   Jefferson Leatherwood Webber (1993)   &   .  \\ 
  Odontoceti   &   Ziphiidae   &   Hyperoodon ampullatus   &   5800   &   Smith etal (2003)   &   .  \\ 
  Odontoceti   &   Ziphiidae   &   Hyperoodon planifrons   &   3000   &   Smith etal (2003)   &   .  \\ 
  Odontoceti   &   Ziphiidae   &   Indopacetus pacificus   &   2200   &   Smith etal (2003)   &   .  \\ 
  Odontoceti   &   Ziphiidae   &   Mesoplodon bidens   &   3400   &   Smith etal (2003)   &   .  \\ 
  Odontoceti   &   Ziphiidae   &   Mesoplodon bowdoini   &   2600   &   Smith etal (2003)   &   .  \\ 
  Odontoceti   &   Ziphiidae   &   Mesoplodon carlhubbsi   &   1400   &   Jefferson Leatherwood Webber (1993)   &   .  \\ 
  Odontoceti   &   Ziphiidae   &   Mesoplodon carlhubbsi   &   3400   &   Smith etal (2003)   &   .  \\ 
  Odontoceti   &   Ziphiidae   &   Mesoplodon carlhubbsi   &   500   &   Mean Walker Houck 1982   &   .  \\ 
  Odontoceti   &   Ziphiidae   &   Mesoplodon densirostris   &   2300   &   Smith etal (2003)   &   .  \\ 
  Odontoceti   &   Ziphiidae   &   Mesoplodon densirostris   &   1033   &   Jefferson Leatherwood Webber (1993)   &   .  \\ 
  Odontoceti   &   Ziphiidae   &   Mesoplodon europaeus   &   5600   &   Smith etal (2003)   &   .  \\ 
  Odontoceti   &   Ziphiidae   &   Mesoplodon europaeus   &   1200   &   Jefferson Leatherwood Webber (1993)   &   .  \\ 
  Odontoceti   &   Ziphiidae   &   Mesoplodon ginkgodens   &   1500   &   Smith etal (2003)   &   .  \\ 
  Odontoceti   &   Ziphiidae   &   Mesoplodon grayi   &   2900   &   Smith etal (2003)   &   .  \\ 
  Odontoceti   &   Ziphiidae   &   Mesoplodon grayi   &   1100   &   Jefferson Leatherwood Webber (1993)   &   .  \\ 
  Odontoceti   &   Ziphiidae   &   Mesoplodon hectori   &   1000   &   Smith etal (2003)   &   .  \\ 
  Odontoceti   &   Ziphiidae   &   Mesoplodon layardii   &   1500   &   Smith etal (2003)   &   .  \\ 
  Odontoceti   &   Ziphiidae   &   Mesoplodon mirus   &   2100   &   Smith etal (2003)   &   .  \\ 
  Odontoceti   &   Ziphiidae   &   Mesoplodon mirus   &   1400   &   Jefferson Leatherwood Webber (1993)   &   .  \\ 
  Odontoceti   &   Ziphiidae   &   Mesoplodon mirus   &   1400   &   Culik (2004)   &   .  \\ 
  Odontoceti   &   Ziphiidae   &   Mesoplodon stejnegeri   &   4800   &   Smith etal (2003)   &   .  \\ 
  Odontoceti   &   Ziphiidae   &   Tasmacetus shepherdi   &   2500   &   Smith etal (2003)   &   .  \\ 
  Odontoceti   &   Ziphiidae   &   Ziphius cavirostris   &   4775   &   Smith etal (2003)   &   .  \\ 
  Odontoceti   &   Ziphiidae   &   Ziphius cavirostris   &   3000   &   Jefferson Leatherwood Webber (1993)   &   .
\end{tabular}
}
\caption{Cetacean size estimates (part 2).}
\label{table:2}
\end{table}

\end{document}